\documentclass[twocolumn,prl,aps,showpacs,footinbib,superscriptaddress,longbibliography]{revtex4-1}
\usepackage{graphicx}
\usepackage{xcolor}
\usepackage{hyperref}
\usepackage{amsmath,amssymb,amsfonts,bbold}
\usepackage[normalem]{ulem}
\graphicspath{{figures/}}

\usepackage{natbib}

\definecolor{darkspringgreen}{rgb}{0.09, 0.45, 0.27}
\definecolor{DebianRed}{rgb}{0.84, 0.04, 0.33}
\definecolor{darkpowderblue}{rgb}{0.0, 0.2, 0.6}

\graphicspath{{fig/}}

\begin{document}

\preprint{}
\title{Boosting the STM's spatial and energy resolution with double-functionalized probe tips
}

\author{Artem Odobesko}
	\altaffiliation{Corresponding author, \texttt{artem.odobesko@uni-wuerzburg.de}}
	\affiliation{Physikalisches Institut, Experimentelle Physik II, 
		Julius-Maximilians-Universit\"{a}t W\"{u}rzburg, Am Hubland, 97074 W\"{u}rzburg, Germany}
\author{Raffael~L.~Klees}
	\affiliation{Institut f\"{u}r Theoretische Physik und Astrophysik, 
		Julius-Maximilians-Universit\"{a}t W\"{u}rzburg, Am Hubland, 97074 W\"{u}rzburg, Germany}
    \affiliation{Institute of Physics, University of Augsburg, D-86159 Augsburg, Germany}
\author{Felix Friedrich}
	\affiliation{Physikalisches Institut, Experimentelle Physik II, 
		Julius-Maximilians-Universit\"{a}t W\"{u}rzburg, Am Hubland, 97074 W\"{u}rzburg, Germany}
\author{Ewelina~M.~Hankiewicz}
	\affiliation{Institut f\"{u}r Theoretische Physik und Astrophysik, 
		Julius-Maximilians-Universit\"{a}t W\"{u}rzburg, Am Hubland, 97074 W\"{u}rzburg, Germany}
\author{Matthias Bode} 
	\address{Physikalisches Institut, Experimentelle Physik II, 
	Julius-Maximilians-Universit\"{a}t W\"{u}rzburg, Am Hubland, 97074 W\"{u}rzburg, Germany}	
	\address{Wilhelm Conrad R{\"o}ntgen-Center for Complex Material Systems (RCCM), 
	Julius-Maximilians-Universit\"{a}t W\"{u}rzburg, Am Hubland, 97074 W\"{u}rzburg, Germany}   %
\date{\today}

\begin{abstract}

Scattering of superconducting pairs by magnetic impurities on a superconducting surface 
leads to pairs of sharp in-gap resonances, known as Yu-Shiba-Rusinov (YSR) bound states.
Similarly to the interference of itinerant electrons scattered by defects in normal metals, 
these resonances reveal a periodic texture around the magnetic impurity. 
However, the wavelength of these resonances is often too short to be resolved even by methods capable of atomic resolution, like scanning tunneling microscopy (STM).
Here, we combine a CO molecule with a superconducting cluster pre-attached to an STM tip to maximize both spatial and energy resolution. 
The superior properties of such a double-functionalized probe are demonstrated 
by imaging the interference of YSR states around magnetic Fe dimers on a Nb(110) surface.
Our novel approach reveals rich interference patterns of both the even and odd combinations of the hybridized YSR states, 
which are inaccessible with conventional STM probes.

\end{abstract}

\maketitle

\begin{figure}[b]
    \includegraphics[width=0.60\columnwidth]{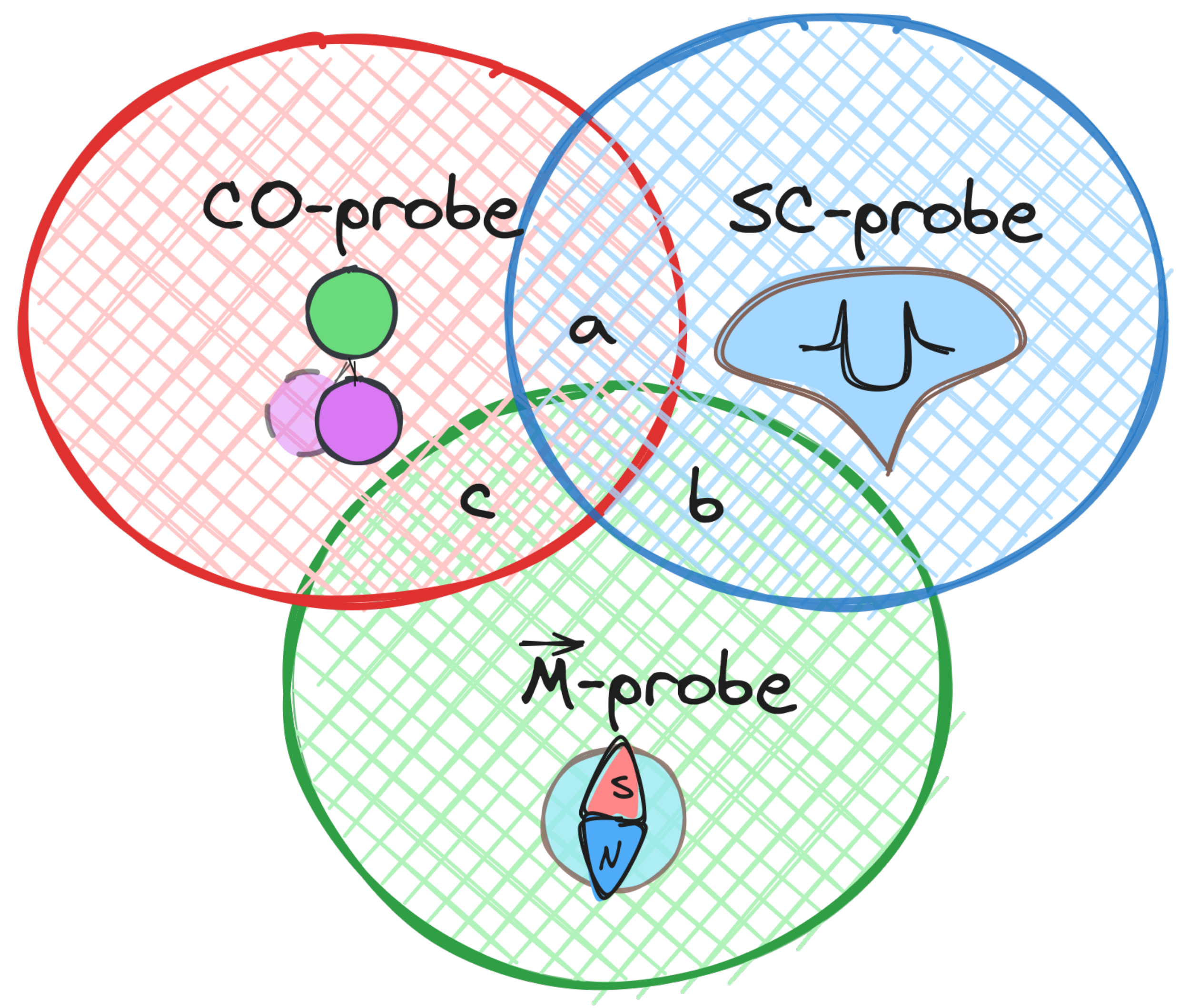}
	\caption{\textbf{Diagram displaying the three basic methods of STM, STS, and SP-STM.} Their sensitivity can be enhanced by functionalization with a CO molecule, a superconducting cluster, or a magnetic atom, resulting in higher spatial resolution in topography, improved energy resolution in spectroscopy, 
		and increased magnetic sensitivity in spin-polarized measurements, respectively. 
		The intersection between these areas represents double-functionalization approaches.}
	\label{fig_1}
\end{figure}

The invention of the scanning tunneling microscope has revolutionized 
our understanding of materials and their properties \cite{Besenbacher1996}.
This progress was made possible by the capability of correlating topographic data of the sample structure 
obtained by constant-current or constant-height scanning tunneling microscopy (STM) \cite{Tersoff1983} 
with the data obtained by scanning tunneling spectroscopy (STS) or spin-polarized (SP)-STM.  
While the former is sensitive to the local density of states \cite{Tersoff1985}, 
the latter grants access to the atomic scale spin structure \cite{Bode2003}. 
However, when performed with normal metal tips, all these methods have their specific limitations, 
which can be overcome by purposive functionalization.  
The spatial resolution of topographic STM measurements can be enhanced 
by attaching a CO molecule to the apex of the STM probe \cite{Gross2009, Weiss2009}, see red circle in Fig.~\ref{fig_1}. 
A superconducting probe boosts the energy resolution in STS beyond the thermal broadening limit \cite{Pan1998} (blue circle) 
and a magnetic atom at the probe apex acts as a spin sensor in SP-STM measurements \cite{Loth2010} (green).
Hereby, the advantages of the probe functionalization not only lie in the improved STM and STS performance, 
but also in the fact that the probe can sequentially be prepared in a single experimental run 
by dressing the apex \textit{in-situ} for specific needs \cite{Friedrich2021}. 

Remarkably, enhancing probe functionalization methods can yield further advancements. The depicted intersection areas \textit{\textbf{a}--\textbf{b}--\textbf{c}} in Fig.~\ref{fig_1} illustrate the potential of double-functionalization in harnessing multiple advantages simultaneously. For example, Ref.~\cite{Schneider2021, Kuester2023} demonstrated the efficacy of combining a magnetic atom with a superconducting probe, significantly enhancing spin contrast at the atomic level (intersection \textit{\textbf{b}} in Fig.~\ref{fig_1}) compared to bulk magnetic tips. Similarly, region \textit{\textbf{c}} involves combinations with magnetic moment-bearing molecules like nickelocene \cite{Czap2019, Mier2021}, resulting in nearly 100\% spin polarization and improved spatial resolution. Yet, the successful double-functionalization to cover intersection area \textit{\textbf{a}} remains elusive. To address this gap, we employ a combination of a superconducting probe and CO molecule, creating a double-functionalized CO-SC-probe. Leveraging improved spectroscopic and spatial resolution, we explore previously inaccessible details in the local density of states (LDOS) around magnetic Fe atoms on a superconducting Nb(110) surface.
A single magnetic impurity results in a pair of particle-hole-symmetric sub-gap resonances, known as Yu-Shiba-Rusinov (YSR) bound states. The wave function of these YSR states (i) reflects the shape of the atomic orbital responsible for magnetic scattering, (ii) oscillates with the Fermi wave vector $k_{\rm F}$ and (iii) decays with the distance $r$ from the impurity \cite{Yu1965, Shiba1968, Rusinov1968}. The decay involves two length scales: an exponential term $\propto e^{-r/\xi_0}$, governed by the superconducting coherence length $\xi_0$, and an algebraic term $\propto(k_{\rm F}\cdot r)^{(1-D)/2}$ that depends on the dimension $D$ of the system. In three-dimensional materials, the wave function diminishes rapidly, primarily due to the algebraic term. This challenge persists for most elemental superconductors as the coherence length $\xi_0$  is much larger than the Fermi wavelength $k_{\rm F}^{-1}$, thereby impeding direct observation of YSR wave function oscillations. Although exclusive observation of such behavior has been successful in lower-dimensional systems or those exhibiting a strong Fermi surface nesting effect \cite{Menard2015, Liebhaber2020, Ruby2016, Kim2020, Ortuzar2022}, it remain elusive in many cases. Moreover, even in the case of magnetic dimers, where more pronounced oscilatory behavior is expected due to the interference of YSR wave functions from individual magnetic atoms \cite{Flatte2000, Morr2003}, numerous STS experiments on various superconducting substrates solely detect broadened initial peaks of odd and even combinations of YSR wave functions \cite{Ji2008, Ruby2018, Meng2015, Kezilebieke2018, Choi2018, Kim2015, Ding2021, Kuester2021a, Beck2021, Friedrich2021}. The challenge of achieving simultaneous high spectroscopic and spatial resolution remains a limiting factor for accessing long-range oscillatory YSR interference patterns.

In this work, by utilizing an innovative double-functionalized CO-SC-probe, we detected unique interference patterns in spatially resolved differential conductance maps of the hybridized YSR states of Fe dimers. Comparing data obtained with and without additional functionalization with a CO molecule, we demonstrate a simultaneous enhancement in spatial and energy resolutions. The distinctive features in the interference maps reveal information about the anisotropy of the Fermi surface of the Nb(110), which are exclusively observed when employing the double-functionalized CO-SC-probe. An analytical model with an anisotropic Fermi contour reproduces the observed interference patterns by introducing a Fermi wave vector  $k_{\rm F} = (9.4 \pm 1.5) \, \rm{nm}^{-1}$.

\subsection{Results}
\begin{figure*}[t]
    \includegraphics[width=0.95\textwidth]{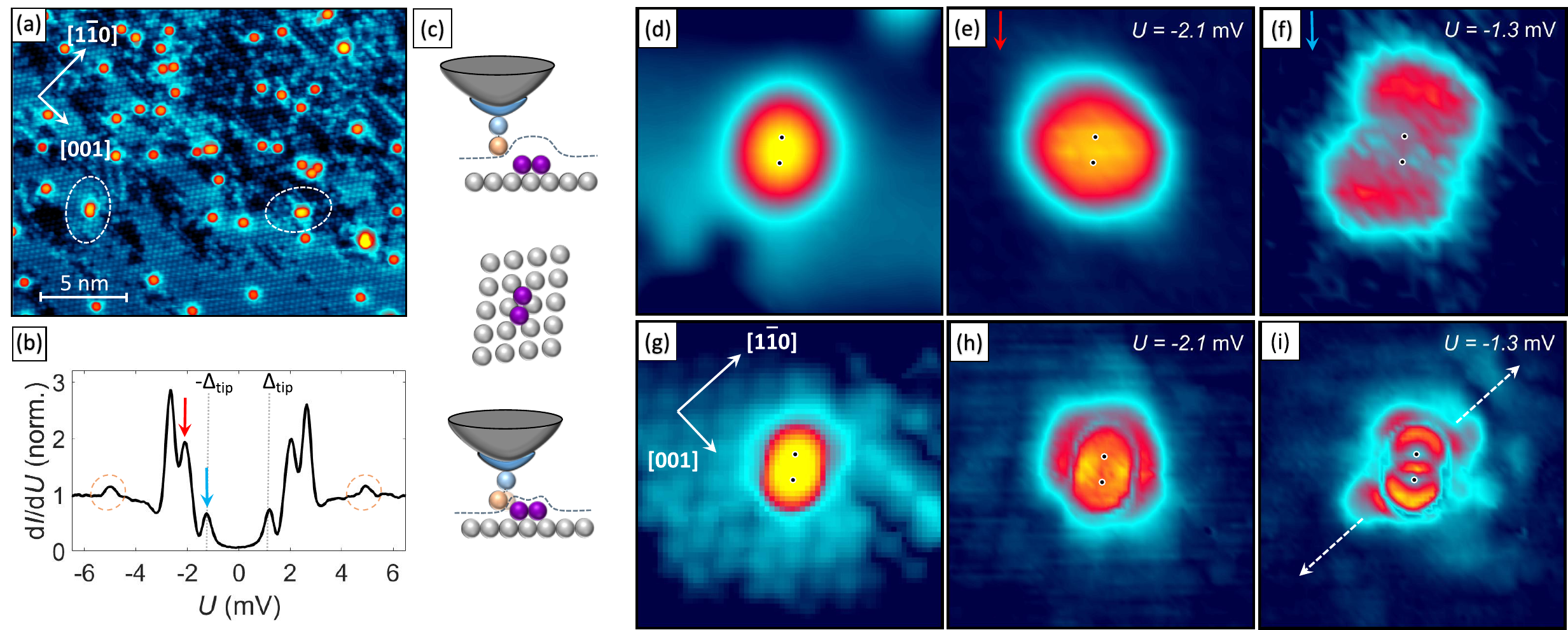}
	\caption{\textbf{Fe atoms on Nb surface.}
        (\textbf{a}) STM topography of Fe atoms (bright protrusions) deposited on Nb(110) captured with a CO-SC probe. Two dimers along the $[1\bar{1}\bar{1}]$ (left) and the $[1\bar{1}1]$ (right), highlighted with dashed ellipses, can be seen. $U_{\rm set} = 10\,{\rm mV}$, $I_{\rm set} = 2\,{\rm nA}$;
        (\textbf{b}) Single ${\rm d}I/{\rm d}U$ spectrum measured in the center of Fe dimer in panel. $U_{\textrm{set}} = 7\,{\rm mV}$, $I_{\rm set} = 0.4\,{\rm nA}$;
        (\textbf{c}) The sketches represent a CO-SC-probe scanning over a magnetic Fe dimer in both non-contact tunneling regime and soft-contact tunneling regime, undergoing Pauli repulsion.
        (\textbf{d}) STM topography of Fe dimer along the $[1\bar{1}\bar{1}]$, scale $3 \times 3 \, {\rm nm}^2$ measured in non-contact tunneling regime. $U_{\rm set} = 7\,{\rm mV}$, $I_{\rm set} = 0.4\,{\rm nA}$;
        (\textbf{e}) Simultaneously measured spatial ${\rm d}I/{\rm d}U$ map at the tunneling bias corresponding to the high-energy YSR state with even symmetry. 	(\textbf{f}) ${\rm d}I/{\rm d}U$ map at the tunneling bias corresponding to the low-energy YSR state with odd-symmetry. 	
        (\textbf{g-i}) Same as in \textbf{d-f} but with CO-SC-probe in a soft-contact mode.  $U_{\rm set} = 7\,{\rm mV}$, $I_{\rm set} = 1\,{\rm nA}$.
        }
	\label{fig_2}
\end{figure*}

Figure~\ref{fig_2}(a) shows a constant-current STM image of Fe atoms deposited on a clean Nb(110) surface, 
taken with the double-functionalized CO-SC-probe. The Nb(110) surface with its lattice constant of $a_{\rm Nb} = 3.3\,\text{\AA}$ is atomically resolved. Dark areas in Fig.~\ref{fig_2}(a) are contaminated with hydrogen or oxygen. The Fe adatoms, visible as bright protrusions, adsorb in four-fold hollow sites of the Nb(110) lattice \cite{Kuester2021, Odobesko2020}. Some Fe atoms spontaneously form dimers.
We will focus on those Fe dimers with the shortest interatomic distance, oriented roughly along the $[1\bar{1}\bar{1}]$ and $[1\bar{1}1]$ directions, equivalent due to surface mirror symmetry. These Fe dimers exhibit energy-split YSR states, as demonstrated in previous studies \cite{Friedrich2021}.

Figure~\ref{fig_2}(b) displays the ${\rm d}I/{\rm d}U$ spectrum measured at the dimer center with a CO-SC-probe, revealing two pairs of YSR states within the energy gap, located at $U = \pm 2.1$~mV and $U = \pm 1.3$~mV. These split resonances arise from the hybridization of single YSR state of the individual Fe atoms \cite{Friedrich2021}. Additionally, distinct peaks at $U = \pm 2.5$mV align with the coherence peaks of the Nb substrate, indicating an additional tunneling path into the substrate. Furthermore, weak resonances at $U = \pm 5.3,{\rm mV}$ are observed, which are absent without a CO molecule at the probe apex. These correspond to the first vibrational mode of the CO molecule, which---for resonant tunneling in SIS junctions---appears as peaks rather than steps in ${\rm d}I/{\rm d}U$ signal, 
as is the case for normal-metallic tips \cite{Homberg2022}.

High-resolution STM imaging with CO-functionalized probes is achieved by leveraging the bending of the molecule at the probe apex due to Pauli repulsion \cite{Weiss2010, Hapala2014, Temirov2015}. At close probe-sample distances, the molecule at the apex undergoes a significant relaxation toward local minima in the interaction potential. The relaxation causes discontinuities in both the frequency shift and tunneling current signal and becomes observable in AFM and STM images as sharp contrast features. 

In Figs.~\ref{fig_2}(d-i), we present two sets of data measured with the CO-SC-probe in a constant-current mode at various probe-sample distances above the Fe dimer. The upper row represents the non-contact regime at low $I_{\text{set}} = 0.4\,{\rm nA}$, while the bottom row illustrates the soft-contact regime at higher $I_{\text{set}} = 1\,{\rm nA}$. 

In case of the non-contact tunneling regime, a close-up of the topography in Fig.~\ref{fig_2}(d) reveals the Fe dimer in the $[1\bar{1}\bar{1}]$ direction. The dimer appears as a single protrusion with a subtle elongation along its axis. The atomic structure of the Nb(110) surface remains elusive, all in one suggesting a low-resolution regime. The black dots mark the positions of the Fe atoms. Figures~\ref{fig_2}(e) and \ref{fig_2}(f) showcase simultaneously measured differential tunneling conductance maps, at tunneling energies aligned with the positions of high- and low-energy hybridized YSR states, respectively. Despite employing a CO-SC-probe, in the non-contact regime the resolution enhancement in ${\rm d}I/{\rm d}U$ maps is minimal, comparable to data obtained with a single-functionalized SC-probe presented in  Ref.~\cite{Friedrich2021}. In particular, any long-range oscillatory YSR interference pattern is absent. Confirming previous observations, the high-energy YSR state at $U = -2.1$\,mV displays an even symmetry, with the strongest signal centered around the dimer. Conversely, the low-energy YSR state at $U = -1.3$\,mV exhibits an odd symmetry, featuring two lobes offset from the dimer center. The replicas at positive tunneling voltages are identical {(see Fig. 3S in Ref.\cite{SupplMat})}.

The topography of the same area at a closer tip-sample distance is shown in Fig.~\ref{fig_2}(g). A direct comparison between the two datasets reveals a significant improvement in spatial resolution, as the atomic resolution of the Nb surface now becomes visible, and individual Fe atoms forming the dimer are distinguishable. Strikingly, a similar enhancement of the spatial resolution  is observed in the differential tunneling conductance maps. Even a superficial inspection of the experimental ${\rm d}I/{\rm d}U$ maps reveals interference patterns which carry a much higher degree of detail compared to the data presented in (e) and (f).

The ${\rm d}I/{\rm d}U$ map of the high-energy YSR state in Fig.~\ref{fig_2}(h) reveals four distinguishable maxima periodically arranged in a direction perpendicular to the dimer axis. These maxima exhibit an elongated shape with the two central ones partially overlapping. This spatial arrangement aligns with the expected even symmetry shown in panel (e). A very different interference pattern is observed for the low-energy YSR state, shown in Fig.~\ref{fig_2}(i), revealing a sequence of arc-shaped maxima arranged along the dimer axis. Up to three maxima are observable on each side, and their intensity rapidly attenuates with increasing distance from the dimer. The arrangement of these maxima strongly suggests an odd symmetry, with a notable exception — the maximum in the center. In the case of an antisymmetric combination of the wave functions, one would anticipate a zero-signal at the center of the dimer. However, the data clearly reveal a sharp peak along the nodal plane of the dimer, in contrast to the data presented in panel (f).


The unexpected positioning of the central maximum in the ${\rm d}I/{\rm d}U$ map of Fig.~\ref{fig_2}(i) arises from the intricate interdependence of the tunneling conductance, which is not solely determined by the convolution of the LDOS of the tip and sample but is also influenced by the tunneling matrix elements between different orbitals. 
Hence, the significance of the tip tunneling orbital cannot be overstated. An accurate representation of the intrinsic spatial distribution of the sample's wave function in ${\rm d}I/{\rm d}U$ maps is only achieved when using an STM tip governed by either $s$- or $p_z$-like frontier orbital, thereby maintaining an even \textit{s}-type symmetry. Nevertheless, it is essential to acknowledge that CO-terminated tips, especially when the CO molecule is bent, induce tunneling through the $p_x$ and $p_y$ orbitals as well, carrying \textit{p}-type odd symmetry \cite{Paulsson2008}.
As discussed in Refs.\cite{Chen1990a, Chen1990b, Chen1991} and later shown here in the context of high-resolution imaging with CO-functionalized probe~\cite{Gross2011}, these frontier \textit{p}-type orbitals results in so-called derivative rules of the tunneling conductance maps, resulting in a contrast inversion of the intrinsic surface LDOS.  Hence, both effects contribute to the measurable spatial ${\rm d}I/{\rm d}U$ maps and should be appropriately taken into account.

%
        
Additionally, closer inspection of the data presented in Fig.~\ref{fig_2}(i) (see also Fig.~S3 in the Supplemental Materials for a mirror symetric dimer in $[1\bar{1}1]$ direction) indicates that the intensity along the arcs is not uniform, but more concentrated along the direction marked by the arrows, where it experiences a weaker attenuation with distance. Comparison with the crystallographic axes reveals that---irrespective of the dimer orientation---this direction is aligned with the Nb $\langle 1\bar{1}0 \rangle$ axis. This observation strongly suggests the presence of a Fermi surface nesting with parallel flat segments of the constant-energy contour along $[1\bar{1}0]$ which provide multiple scattering vectors, resulting in the the so-called ``focusing effect'' with a longer propagation of the YSR along this specific direction \cite{Ruby2016,Kim2020,Ortuzar2022}.

\begin{figure}[t]
    \includegraphics[width=1\columnwidth]{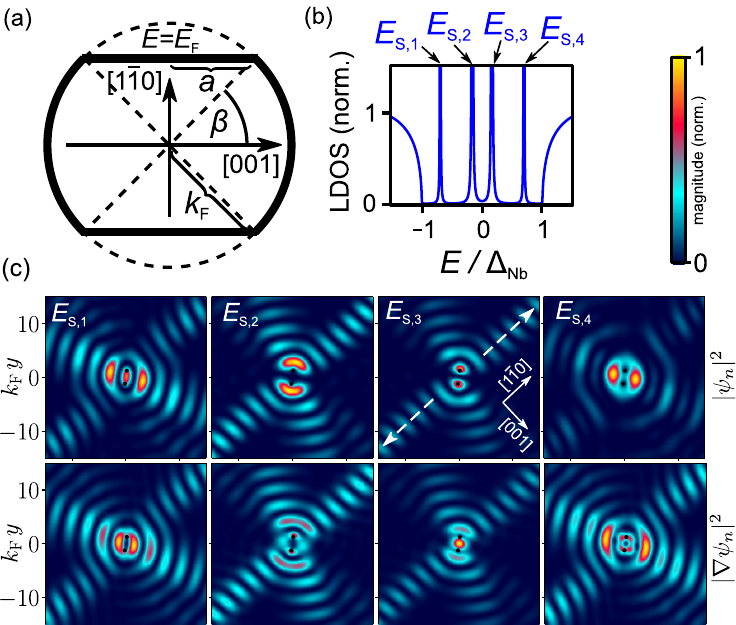}
	\caption{
    \textbf{Fermi surface model and resulting DOS.} 
    (\textbf{a}) 
        Sketch of the stadium-shaped Fermi surface at the Fermi energy $E_{\rm F}$. $a \in [0,k_{\rm F}]$ and the angle $\beta = \arccos(a / k_{\rm F})$ define the regions with flat segments, where $a = 0$ corresponds to a circular Fermi surface with radius $k_{\rm F} > 0$ and $\beta = \pi /2$.
	(\textbf{b}) 
        LDOS at the impurity position $\boldsymbol{r} = \boldsymbol{r}_1$ normalized to its large-energy value at $E = 10^3 \Delta_{\rm{Nb}}$ for a pair of magnetic impurities located at $\boldsymbol{r}_{1,2}$. There are two pairs of YSR states with energies $E_{\rm S,4} = -E_{\rm S,1} \approx \pm 0.69 \, \Delta_{\rm{Nb}}$ and $E_{\rm S,3} = -E_{\rm S,2} \approx \pm 0.15 \, \Delta_{\rm{Nb}}$.
    (\textbf{c})
        Surface LDOS $|\psi_n|^2$ (top row), gradient $|\nabla\psi_n|^2$ (middle row). The plots are normalized to their maximal value. Black dots represent the locations of the individual magnetic impurities.
    	Data were rotated to fit the orientation in Fig.~\ref{fig_2}.
	Parameters for (\textbf{b}) and (\textbf{c}):  $a = 0.6\, k_{\rm F} $, $\xi_0 = 100/k_{\rm F}$, $k_{\rm F} = 2.5/d$, $J = -0.82/N_0$, $U = 0.70/N_0$, 
	where $N_0$ is the normal-state density of states at the Fermi energy.
    }      
    \label{fig_3}
\end{figure}

\subsection{Theory}
To rationalise our results, we model the experimental data with a simplified 2D continuous model of a bare superconducting surface with an anisotropic stadium-shaped Fermi surface, as shown in Fig.~\ref{fig_3}(a). A lattice model with a similar Fermi surface was already successfully used in Ref.~\cite{Odobesko2020a}. We introduce a parameter $a \in [0,k_{\rm F}]$ that defines the length of the flat segments of the Fermi contour, where $a = 0$ corresponds to a circular shape with the Fermi wave vector $k_{\rm F}$. 
Since the Fermi velocity points always perpendicular to the Fermi surface contour, $a > 0$ generates a focusing of the hybridized YSR-state propagation along the $[1\bar{1}0]$-direction in real space, whose strength increases with increasing $a$. 
In the following, we focus our analysis on the LDOS of the impurity-dressed substrate, keeping in mind that the experimental data obtained from STS is actually a convolution of the LDOS of the substrate and the tip. The influence of the tip orbital will be discussed further.

Following the Green's function approach described in Ref.~\cite{Ortuzar2022}, we model the Fe dimer by adding two magnetic impurities at an inter-impurity distance $d$ on the bare superconducting substrate. We fix $d = (2.66 \pm 0.18)\,\text{\AA}$ to the average between the two extremes Fe and Nb with nearest-neighbor distances of $2.48\,\text{\AA}$ and $2.85\,\text{\AA}$, as discussed in the Supplemental Material \cite{SupplMat}.
We also account for the epitaxial strain between the Fe dimer and the Nb(110) substrate, where the axis of the Fe dimer is rotated clockwise by an angle of $4^\circ$ with respect to the $[1\bar{1}1]$ direction \cite{SupplMat}.
For simplicity, these impurities are assumed to be identical and described by a semiclassical Shiba model \cite{Shiba1968} with an onsite energy $U$ and an exchange coupling $J$.  

The Fermi momentum $k_{\rm F}$ is chosen to match the oscillation pattern in the experimental data.
We find a good match between our theoretical model with a simplified shape of the Fermi contour 
and the experimental data for $k_{\rm F} = 2.5/d \approx (9.4 \pm 1.5) \, \text{nm}^{-1}$, see Fig.~S5 in \cite{SupplMat}. 
Surprisingly, this value for $k_{\rm F}$ estimated for Fe impurities on Nb(110) is almost two times larger than the one obtained for Mn on Nb(110) \cite{Schneider2022}. 
Yet, it is important to note that $k_{\rm F}$ for Fe atoms is estimated for YSR states which correspond to scattering channels related to the $d_{z^2}$-orbital, whereas for Mn $k_{\rm F}$ is obtained for the $d_{yz}$-orbital \cite{Schneider2022}.
We speculate that the relatively large difference in the effective Fermi wave length $k_{\rm F}$ 
for screening magnetic impurities is caused by the fact that the $d_{yz}$-states of Mn and the $d_{z^2}$-states of Fe hybridize with very different bands of the Nb Fermi surface.  
Further \textit{ab initio} calculations would be desirable to clarify this issue.

In Fig.~\ref{fig_3}(b), we show the resulting LDOS at one of the impurity sites, which shows two pairs of hybridized YSR bound states at the energies $E_{{\rm S},n}$ ($n = 1,2,3,4$) with  $E_{{\rm S},4} = - E_{{\rm S},1} \approx 0.69 \,  \Delta_{\rm{Nb}}$ and $E_{{\rm S},3} = - E_{{\rm S},2} \approx 0.15 \,  \Delta_{\rm{Nb}}$.
The parameters $U$ and $J$ are chosen such that 
the energy difference of the positive and negative pair is $\Delta E = |E_{\rm S,4}|-|E_{\rm S,3}| \approx 0.54\, \Delta_\mathrm{Nb}$, which corresponds to the experimentally observed value $\approx 0.8$meV.

In the first row in Fig.~\ref{fig_3}(c), we show the spatial behavior of the LDOS, i.e., the wave function $|\psi_n|^2$, of the four YSR bound states.
While the high-energy pair of YSR states shows even symmetry with a finite value at the origin, the low-energy pair shows odd symmetry and vanishes at the origin. 
The calculated LDOS maps are in good agreement with the tunneling conductance maps observed experimentally.
In particular, the LDOS map at the energy $E_{{\rm S},1}$ ($E_{{\rm S},2}$) qualitatively reproduces the high (low) LDOS in the nodal plane at for the even- (odd)-symmetric YSR state at $U = -2.1\,\mathrm{mV}$ ($U = -1.3\,\mathrm{mV}$). 
Furthermore, the model reproduces the ``focusing effect'' along the $[1\bar{1}0]$ direction, marked with arrows in Fig.~\ref{fig_3}(c).
Since the theoretical model is two-dimensional, the attenuation of YSR wave function is significantly reduced and an additional set of maxima in the other direction is also observed. 
They arise from the remaining circular segment of the Fermi contour and their direction rotates with the orientation of the dimer, whereas the ``focusing'' direction is independent of the dimer orientation and always directed along $[1\bar{1}0]$.
However, the experimental measurement data in Fig.~\ref{fig_2}(i) clearly shows a peak at the origin, which we attribute to tunneling through $p_{x,y}$-orbitals of the CO molecule at the tip apex. 
To model the measurement of the differential conductance using STM tips with nonisotropic orbitals, i.e., beyond the $s$-orbital, one needs to take into account the shape of the tip orbital wave functions and their spatial overlap with the substrate YSR bound-state wave function.
This results in different transition matrix elements for tunnel processes through different orbitals, which result in so-called derivative rules \cite{Chen1990a,Chen1990b,Chen1991}.
In particular, tunneling through $p_\alpha$-orbitals $(\alpha = x,y,z)$ leads to transition matrix elements proportional to $|\partial_\alpha \psi_n|^2$.
Due to the uncertain and probably stochastically fluctuating azimuthal alignment of the $p_x$- and $p_y$-orbitals of the CO molecule with respect to the Fe dimer in a soft-contact regime, we assume that tunneling occurs equally through both orbitals. Consequently, the only measurable quantity becomes the gradient of the wave function in the $x$-$y$-plane, represented as $|\nabla \psi_n|^2$ and as depicted in the second row of Fig.~\ref{fig_3}(c).
%
Here, the gradients of the odd YSR states show a peak in the center (although less pronounced in the second plot), while the gradients of the YSR states with even symmetry vanish at the origin. 

Moreover, as detailed in the Supplementary Material \cite{SupplMat}, our theoretical framework anticipates an energy-dependent phase shift between electron-like and hole-like YSR states, arising from the counter movement of electrons and holes, observed in Ref.\cite{Menard2015}. This phenomenon is distinctly evident in Fig.~\ref{fig_3}(c), illustrating the difference in spatial distribution between YSR states with positive and negative energies. However, our observed tunnelling conductance maps do not show such a phase variation, likely due to the combined influence of $p_z$ and $p_{x,y}$ orbitals, which mixes and obscures these effects.

In conclusion, our findings clearly demonstrate the advantages of double-functionalized STM probes, comprising a superconducting cluster and an additional CO molecule attached to it, simultaneously maximizing energy resolution and significantly enhancing spatial resolution.
We have shown that in the contact tunneling regime, the bending of the CO molecule at the tip apex leads to an increase in spatial contrast in STS, akin to effects observed in high-resolution STM/AFM microscopy. However, this also results in additional tunneling through $p_{x,y}$ orbitals, which yields a signal proportional to the square of the sample wave function gradient, further enhancing spatial resolution, albeit introducing symmetry alterations in observed features.
Accounting for these effects allows an STM equipped with such a doubly functionalized probe to afford unique access to the wave functions of YSR-bound states in bulk three-dimensional superconductors. This renders such CO-SC-probe a powefull tool for investigating odd-frequency and p-wave superconductivity, exploring magnetic chains with Majorana end states, and other quantum phenomena on a atomic scale.

\subsection{Methods}
The experiments are performed in a home-built low-temperature STM at a base temperature of 1.4\,K. 
The Nb(110) surface is cleaned by a series of high-temperature flashes \cite{Odobesko2019}. 
Fe atoms are deposited \textit{in-situ} onto the Nb substrate at a temperature of 4.2\,K.
To get a superconducting probe, an electro-chemically etched W tip was brought in contact 
with the Nb crystal, thus creating a Nb cluster on the tip apex. 
CO molecules were picked up from a clean Cu(001) surface using the procedure described in Ref.~\cite{Friedrich2021}.
The resulting double-functionalized tips exhibit a superconducting gap $\Delta_\mathrm{tip}$ 
which corresponds to about 70-90\% of the bulk Nb value \cite{Bose2005}.
The presence of a superconducting gap in the LDOS of the tip causes a corresponding shift 
of the sample's LDOS features in the conductance spectra by $\Delta_\mathrm{tip}$.
The experimental data are obtained within the tunneling regime, where tunnel resistances $R_\mathrm{tun} > 10^6\, \Omega$. This ensures that the tunneling current is predominantly governed by single-electron tunneling event rather than Andreev reflections, observed in both non-contact and contact regimes.
 \cite{Ternes2006, Ruby2015, Villas2020, Villas2021}.
All spectroscopic measurements are performed with a modulation voltage of 0.1\,mV at a frequency of 890\,Hz.

\bibliography{bibliography_11}

\subsection{Acknowledgments}
\begin{acknowledgments}
R.L.K.~acknowledges fruitful discussions with Juan-Carlos Cuevas and Daniel Gresta.

\textbf{Funding:} This work was supported by Deutsche Forschungsgemeinschaft (DFG, German Research Foundation) through SFB 1170 (project C02). We also acknowledge financial support by DFG under Germany's Excellence Strategy through W{\"u}rzburg-Dresden Cluster of Excellence on Complexity and Topology in Quantum Matter -- ct.qmat (EXC 2147, project-id 390858490) 

\textbf{Author contributions:}
A.O. and F.F. conducted the experiments and analyzed the data, R.L.K. carried out the simulations and calculations,  A.O. conceived and directed the study,  M.B. and E.M.H. supervised the project, A.O. and R.L.K. wrote the initial version of the manuscript. All authors discussed the results and their interpretation. A.O., R.L.K., F.F. and M.B. contributed to the manuscript.

\textbf{Competing interests:} The authors declare that they have no competing interests.

\textbf{Data and materials availability:} All data needed to evaluate the results and conclusions in the paper are present in the paper and/or the Supplementary Materials.

\end{acknowledgments}

\end{document}